\documentclass[conference]{IEEEtran}
\IEEEoverridecommandlockouts
\usepackage{orcidlink}
\usepackage{cite}
\usepackage{amsmath,amssymb,amsfonts}
\usepackage{algorithmic}
\usepackage{algorithm2e}
\usepackage{graphicx}
\usepackage{textcomp}
\usepackage{xcolor}
\usepackage{csquotes}
\usepackage{listings}
\usepackage[frozencache,cachedir=minted-cache]{minted}
\usepackage{caption}
\usepackage{booktabs}
\usepackage{subcaption}
\usepackage{url}
\usepackage{siunitx}
\usepackage{placeins}
\usepackage{xurl}

\usepackage{todonotes}
\presetkeys{todonotes}{inline}{}

\lstdefinestyle{yaml}{
    frame=single,
    captionpos=b,
    basicstyle=\color{blue}\footnotesize\ttfamily,
    breaklines=true,
    rulecolor=\color{black},
    string=[s]{'}{'},
    stringstyle=\color{blue},
    comment=[l]{:},
    commentstyle=\color{black},
    morecomment=[l]{-}
 }

\makeatletter
\newcommand{\linebreakand}{%
  \end{@IEEEauthorhalign}
  \hfill\mbox{}\par
  \mbox{}\hfill\begin{@IEEEauthorhalign}
}
\makeatother

\def\BibTeX{{\rm B\kern-.05em{\sc i\kern-.025em b}\kern-.08em
    T\kern-.1667em\lower.7ex\hbox{E}\kern-.125emX}}

\DeclareCaptionType{equ}[][]

\newcommand*\dottvar[1]{\ifx\relax#1\else
  \expandafter\ifx\string_#1\string_\allowbreak\else#1\fi
  \expandafter\dottvar\fi}
  
\newcommand{\ttt}[1]{\texttt{\expandafter\dottvar\detokenize{#1}\relax}}

\def\tablename{Table}

\begin{document}

\title{Closing the HPC-Cloud Convergence Gap: Multi-Tenant Slingshot RDMA for Kubernetes
\thanks{
This research is supported by the European Commission under the Horizon project OpenCUBE (101092984).
This work has been accepted for publication in IEEE Cluster 2025.

© 2025 IEEE. Personal use of this material is permitted. Permission
from IEEE must be obtained for all other uses, in any current or future
media, including reprinting/republishing this material for advertising or
promotional purposes, creating new collective works, for resale or
redistribution to servers or lists, or reuse of any copyrighted
component of this work in other works.
}
}

\author{
\IEEEauthorblockN{1\textsuperscript{st} Philipp A. Friese} \\
\IEEEauthorblockA{\textit{Technical University of Munich} \\
    Garching, Germany \\
    philipp.friese@cit.tum.de\,\orcidlink{0000-0002-3124-5364}} \\
\and
\IEEEauthorblockN{2\textsuperscript{nd} Ahmed Eleliemy} \\
\IEEEauthorblockA{\textit{HPE HPC/AI EMEA Research Lab} \\
    Basel, BS, Switzerland \\
    ahmed.eleliemy@hpe.com\,\orcidlink{0000-0003-3258-1738}} \\
\linebreakand
\IEEEauthorblockN{3\textsuperscript{rd} Utz-Uwe Haus} \\
\IEEEauthorblockA{\textit{HPE HPC/AI EMEA Research Lab} \\
    Wallisellen, ZH, Switzerland \\
    uhaus@hpe.com\,\orcidlink{0000-0001-7292-9984}} \\
\and
\IEEEauthorblockN{4\textsuperscript{th} Martin Schulz} \\
\IEEEauthorblockA{\textit{Technical University of Munich} \\
    Garching, Germany \\
    schulzm@in.tum.de\,\orcidlink{0000-0001-9013-435X}}
}

\maketitle


\begin{abstract}
Converged HPC-Cloud computing is an emerging computing paradigm that aims to support increasingly complex and multi-tenant scientific workflows.
These systems require reconciliation of the isolation requirements of native cloud workloads and the performance demands of HPC applications.
In this context, networking hardware is a critical boundary component: it is the conduit for high-throughput, low-latency communication and enables isolation across tenants.
HPE Slingshot is a high-speed network interconnect that provides up to 200~Gbps of throughput per port and targets high-performance computing~(HPC) systems.
The Slingshot host software, including hardware drivers and network middleware libraries, is designed to meet HPC deployments, which predominantly use single-tenant access modes.
Hence, the Slingshot stack is not suited for secure use in multi-tenant deployments, such as converged HPC-Cloud deployments.
In this paper, we design and implement an extension to the Slingshot stack targeting converged deployments on the basis of Kubernetes.
Our integration provides secure, container-granular, and multi-tenant access to Slingshot RDMA networking capabilities at minimal overhead.
\end{abstract}

\begin{IEEEkeywords}
HPE Slingshot, Kubernetes, RDMA, Converged HPC-Cloud
\end{IEEEkeywords}

\section{Introduction}
\label{sec:introduction}

Network communication is a key factor in the overall performance of classic High Performance Computing (HPC) workloads, which have strict bandwidth and latency requirements and usually utilize Remote Direct Memory Access~(RDMA) to achieve optimal communication performance. 
HPC systems often deploy dedicated network hardware explicitly designed to meet these communication requirements. 
The performance characteristics of HPC network hardware, however, are becoming increasingly relevant not just for classic HPC deployments, but also for other computing paradigms such as converged HPC-Cloud deployments.

Converged HPC-Cloud computing is an emerging computing paradigm that caters to the requirements of complex, scientific workflows using a multi-tenant deployment model.
These converged systems must reconcile the tenant isolation requirements of native cloud workloads and the minimal-overhead requirements of HPC workloads.

HPE~Slingshot is a high-speed, Remote Direct Memory Access~(RDMA) interconnect that provides up to \qty{200}{Gbps} of throughput per port, with the upcoming generation reported to provide up to \qty{400}{Gbps}~\cite{sh400}. It is designed for HPC systems and is used by a large number of leading supercomputers. 
Given its design-focus, however, Slingshot caters to the requirements and deployment models of HPC systems, which are often incompatible with those of converged HPC-Cloud systems. 
In particular, these systems require strong confidentiality and user isolation, typically achieved by virtualizing both compute and network resources. 
This is implemented, for example, using containerization for compute resources and overlay networks for network resources.
These requirements are not met by the HPE Slingshot network hardware, making it incompatible with converged HPC-Cloud deployments.

We aim to close this identified gap by introducing a novel integration of HPE Slingshot into the software stack of converged HPC-Cloud systems. 
Specifically, we extend HPE Slingshot to support secure, multi-tenant RDMA-based communication for workflows running in the container orchestrator Kubernetes~\cite{k8s}. 
For this, we extend the Slingshot host software to support containers and implement a Container Network Interface (CNI) plugin. 
We develop a dedicated service for managing the lifetime of Slingshot Virtual Networks and their Virtual Network IDs (VNIs), which provide network isolation between communication parties, and integrate this isolation domain via a custom resource into Kubernetes.
With our integration of Slingshot into cloud-native software stacks, we aim to further enable converged HPC-Cloud cluster systems. 

We envision two classes of use-cases of our integration: (1) user-level co-located applications, and (2) system-level co-located applications. Use-case (1) targets co-scheduled applications, which should not be able to interfere with each other and which may benefit from using different traffic classes. This may be desirable for example when co-scheduling a low-latency critical application with a less latency-sensitive task such as check-pointing. Another scenario for use-case (1) are multi-tenancy deployments, where individual applications must not be able to access or interfere with the network traffic of other applications from potentially different users.
Use-case (2) targets co-scheduling of system-level applications next to user applications, such as monitoring or administration tools, where interference is similarly undesirable.

In this paper, we make the following contributions: 

\begin{itemize}
    \item We extend the Slingshot driver and library with support for network-namespace based authentication and introduce a CNI plugin, which manages container-granular access to Slingshot network hardware.
    \item We develop a VNI Service, which integrates Slingshot VNIs as a resource into Kubernetes clusters to provide secure, isolated RDMA communication between containers.
    \item Finally, we present an overhead evaluation of our introduced software stack, both in terms of network communication and job admission overhead.
\end{itemize}

The remainder of this paper is structured as follows: 
In Section~\ref{sec:background} we briefly introduce the background on RDMA, HPE Slingshot, and container networking. 
In Section~\ref{sec:implementation} we describe our implementation in detail, covering the extension of the Slingshot driver and library, our CNI plugin, as well as the VNI service.
In Section~\ref{sec:evaluation}, we evaluate the overhead of our integration.
In Section~\ref{sec:related-work} we describe related work, and in Section~\ref{sec:conclusion} we provide concluding remarks.

\section{Background}\label{sec:background}

In this section, we briefly introduce the background on Remote Direct Memory Access, the HPE Slingshot network hardware and accompanying software stack, as well as container networking.

\subsection{Remote Direct Memory Access (RDMA)}

Remote Direct Memory Access (RDMA) enables applications to perform direct memory access operations across nodes without invoking kernel-based network operations, as with standard TCP/IP-based communication. It is the prevalent communication mechanism used in HPC due to its low latency and high throughput communication properties. RDMA-based communication is typically established in two steps: 

First, the application requests the creation of an RDMA endpoint~(EP) from the network interface card~(NIC). 
An RDMA-EP is a handle to control structures and a set of queues, which can be used by the application to enqueue transmission commands and handle data receival. 
Endpoint allocation is typically a secured operation via the kernel driver of the used RDMA NIC and, hence, requires a switch into kernel space. 
The handle to the established EP is returned to the application for future communication.

Second, the application uses the returned EP handle to send packets directly to the NIC. 
Commands posted to the endpoint queues are directly accessible by the NIC. 
Similarly, the application can register memory regions with the NIC for direct memory access.
Communication via RDMA endpoints, therefore, involves reads and writes to a set of registered memory regions and queues, neither of which requires context switches into kernel code.


\subsection{HPE Slingshot}

HPE Slingshot is a high-speed network solution widely used in leading HPC systems. It comprises a network switch, named Rosetta, and the network ASIC or NIC named Cassini 11 (CXI). 
The CXI NIC provides both classical Ethernet-based communication via a regular Linux network device, and RDMA-based communication via a character device.
The RDMA network stack exists outside of classical Linux networking paradigms, such as routing tables or network namespaces, matching the kernel-bypass paradigm of RDMA-based communication.

\subsection{Slingshot Access Model}

The Slingshot network uses two components to govern access of users to the high-speed network: (1) Virtual Networks and (2) CXI Services. These components are analogous to Protection Domains~\cite{rfc5042}, which are used in InfiniBand-based RDMA networks, in that they restrict access to communication resources to a set of authorized users.

\textbf{Virtual Networks} and their associated Virtual Network IDs (VNI), which are represented as unsigned integers, provide layer-2 network isolation domains, similar to VLANs. The Rosetta switch can be configured to strictly enforce VNIs and only route packets within a VNI if both the sender and receiver NIC have been granted access to that VNI. 
\emph{Slingshot Virtual Networks provide a secure isolation domain between NICs}.

\begin{figure}
    \centering
    \includegraphics[width=.70\linewidth]{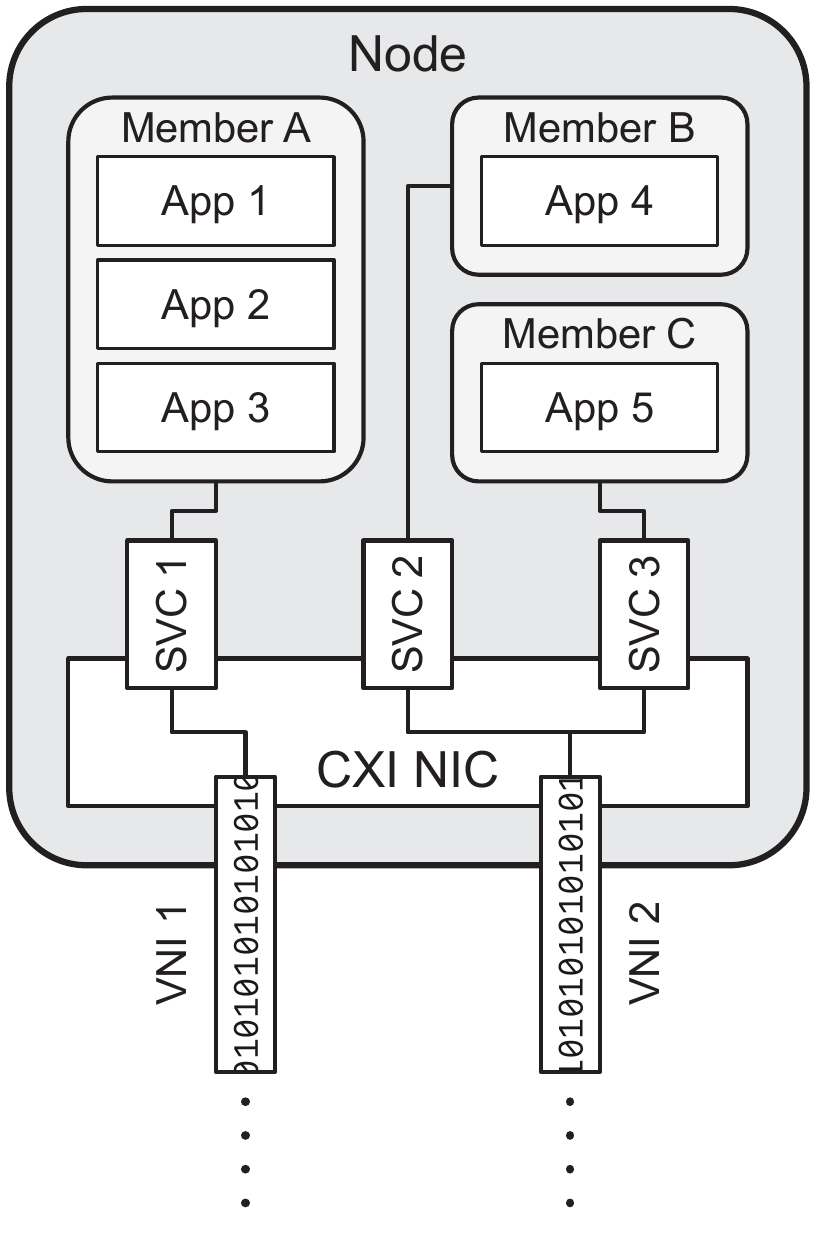}
    \caption{Slingshot Access Model | Virtual Networks (VNIs) are used to establish an isolated network between NICs. CXI services (SVC) are used to grant VNI access to applications of whitelisted members. }
    \label{fig:cxi-vni-svc}
\end{figure}

\textbf{CXI Services} (SVC) are part of the CXI NIC driver and provide user-granular access to NIC resources. They are used to restrict access to a set of VNIs for authorized members and can be configured to limit the use of communication resources, such as transmission or event queues.
\emph{CXI Services provide member-granular access to VNIs}. 
Figure~\ref{fig:cxi-vni-svc} outlines the access model architecture for VNIs and CXI services.

During RDMA endpoint creation, users request access to one or more VNIs via calls to the CXI library~\emph{libcxi}. This library then checks whether any CXI service exists that (1) lists the requesting user as an authorized member, and (2) is authorized to use the requested VNIs.
Members of a CXI service are authenticated via Linux user and group IDs. 

Authentication against CXI services is only performed during endpoint creation. Subsequent communication using the acquired endpoint does not require authentication, which is in line with the kernel bypass functionality of RDMA. 

CXI service configuration requires privileged permissions and is done either ahead of time during user onboarding or dynamically, for example, via a daemon running as root. The latter approach is implemented, for instance, in Slurm via the daemon \ttt{slurmd}, which creates the required services during job creation. Alternatively, the HPE-provided Dynamic RDMA Credential (DRC) 2 mechanism can be used, which allows users to request new VNIs at run time. 
In both cases, VNIs must be assigned mutually exclusively to users to maintain per-user network isolation, which in-turn requires some form of VNI management to maintain this exclusivity.

\subsection{Container Networking}

Linux containers provide operating system-level isolation of resources and are implemented using Linux control groups (\ttt{cgroups}), which control resource usage such as CPU time or memory, and Linux namespaces, which partition and isolate resources, for example, by providing an isolated view on user and group IDs via user namespaces or network components via network namespaces.

Network devices, such as physical NICs, or virtual devices, such as a bridge, can only reside in at most one network namespace. By default, network devices are placed in the host network namespace. Containers are usually started in a new network namespace, which cannot access the host's network devices. To facilitate communication,  virtual Ethernet (\ttt{veth}) device pairs are used. One \ttt{veth} device is attached to the isolated container network namespace, and the other to the target namespace, which can be either the host namespace or the namespace of another container. 
Based on these namespaces, overlay networks can be created, which allow subsets of containers running on multiple nodes in a cluster to communicate in isolation. 
Due to the involvement of virtual components, the performance of overlay networks is usually prohibitive for HPC workloads~\cite{Beltre_Saha_Govindaraju_Younge_Grant_2019,Keller_Tesser_Borin_2023}.

Overlay networks can be created using the Container Network Interface (CNI)~\cite{cni},
which specifies an interface for CNI plugins to create, manage, and destroy overlay network plugins. During container creation, the CNI plugin is called with elevated permissions to, for example, create a new \ttt{veth} pair, configure it with the correct routing and address information, and attach it to the container network namespace. Conversely, the CNI plugin is called during container teardown for interface deletion and potential recycling of network resources.

\section{Design and Implementation}
\label{sec:implementation}

The Slingshot access model relies on the Linux permission model and authenticates VNI users via their user and group IDs. This authentication scheme is insufficient for namespace-based deployments, such as containers in converged HPC-Cloud stacks, for two reasons:

First, to provide kernel-bypass functionality, the CXI NIC exposes its RDMA communication capabilities through a character device, which is not integrated into the Linux kernel's networking paradigms. Classical container networking approaches, such as \ttt{veth} pairs, are therefore not applicable to Slingshot.

Second, user namespaces~\cite{usernamespaces} isolate user and group IDs in a container from those on the host namespace. As a consequence, users can be given root permissions, or UID 0, inside a container, which is mapped to a different and potentially unprivileged UID on the host. Subsequently, users can freely change their UID and GID inside the container and therefore authenticate as any user against any CXI services. 

The CXI driver can be modified to respect user namespaces and therefore use the host user and group ID. However, a follow-up problem arises in case the widely used container orchestrator Kubernetes is used: Kubernetes abstracts identity management from Linux permissions and implements a separate, cluster-wide identity system. As a consequence of this abstraction, all Kubernetes containers are executed by one user. Even a user namespace-aware CXI driver cannot differentiate between two containers and therefore cannot distinguish between different users.

\begin{figure}
    \centering
    \includegraphics[width=.8\linewidth]{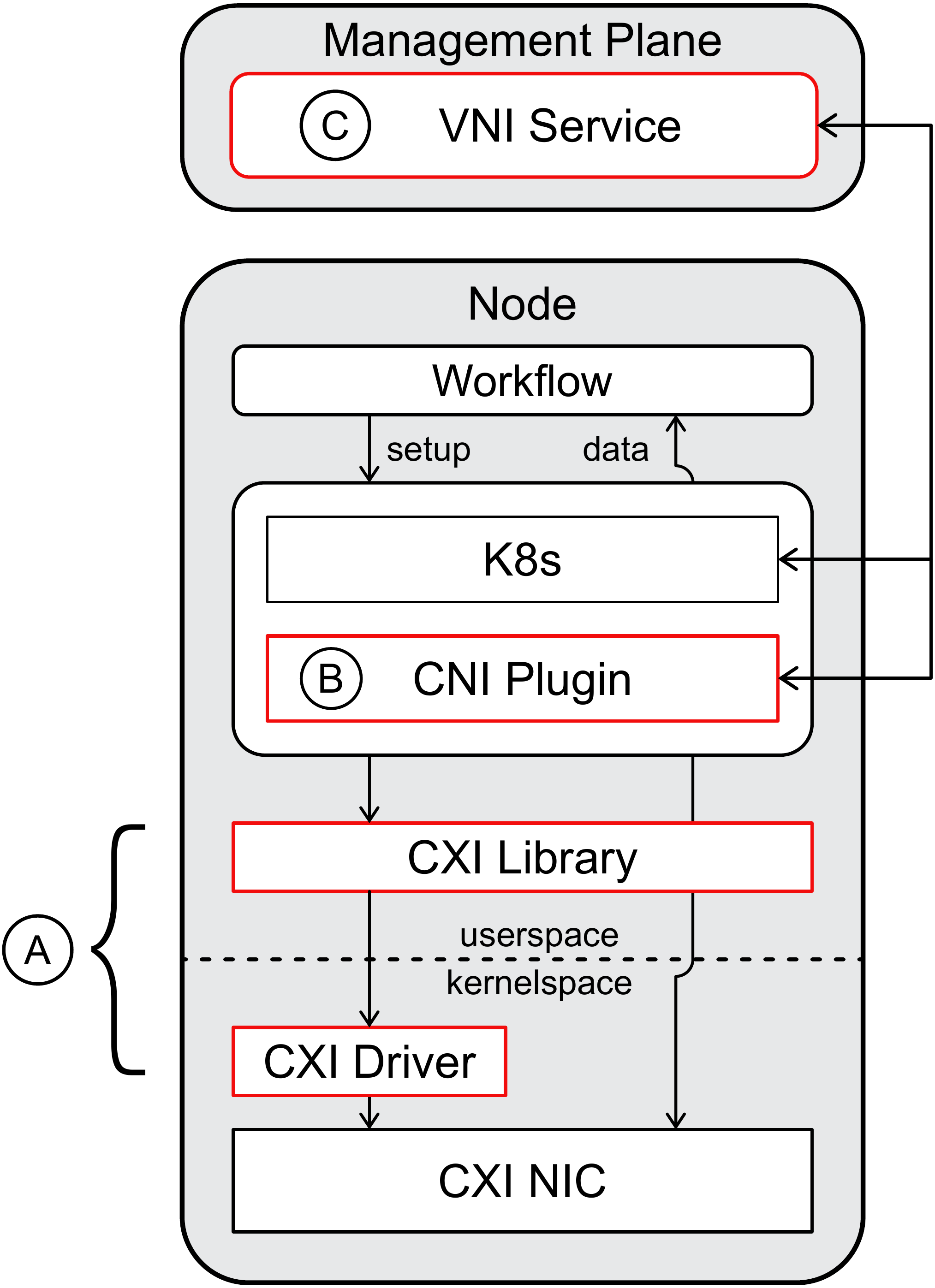}
    \caption{Design of our Slingshot Software Stack | Red components are contributions of this paper.}
    \label{fig:shs-stack}
\end{figure}

We address this problem and outline our design and implementation of a container-granular access model for Slingshot, integrating it into the container orchestrator Kubernetes. We introduce contributions on three levels: (A) the CXI driver and library, (B) a custom CXI CNI plugin, and (C) a Kubernetes-compatible VNI management service. Figure~\ref{fig:shs-stack} presents an overview of our design: red components are contributions of this paper and are discussed in detail in the following sections.

\subsection{CXI Driver and Library}\label{subsec:cxi-driver-library}

User authentication and subsequent authorization to CXI services are implemented in the CXI kernel driver. The current driver checks for a list of members, which are either identified by their UID or GID. During RDMA endpoint creation, the calling user eventually requests a CXI service and calls into the CXI kernel driver code. This code extracts the UID and GID from the calling process and checks whether the requested CXI service contains members with matching UID or GIDs, based on the configured service member type.

As outlined above, this authentication scheme is not compatible with network-namespace-based deployments. We solve this incompatibility by extending the CXI driver with a third member type: the network namespace or \ttt{netns} member. Each network namespace is uniquely identifiable by the network namespace ID, which is stored by the kernel as an unsigned integer. This ID corresponds to the inode of the associated network namespace file and can be retrieved using procfs~\cite{procfs}. 

Authentication occurs similarly to UID or GID-based authentication. When a new RDMA endpoint is requested, the extended CXI kernel driver extracts the network namespace ID from the requesting process via procfs and compares it to the ID configured for the requested CXI service. Since network namespaces are governed outside of application control, malicious users inside a container cannot modify their network namespace ID, unlike their UID and GID. 

We choose network namespaces as a point of authorization because they are the natural isolation mechanism used in containers, which form a large building block of converged HPC-Cloud stacks. 
If two containerized processes do not share network namespaces, then they should not share networking resources. Conversely, two processes sharing one network namespace automatically share all Linux networking resources attached to that namespace. Our design extends this notion to Slingshot RDMA resources.

The addition of the \ttt{netns} service member type is also extended to the userspace library \emph{libcxi} and to the network abstraction library \emph{libfabric}~\cite{libfabric}, which is the de-facto interface for Slingshot.

CXI services are managed outside the lifetime of applications and must be created before a (potentially containerized) application can request an RDMA endpoint. We implement this functionality via a CNI plugin.

\subsection{CXI CNI Plugin}\label{subsec:cxi-cni}

We develop a Container Network Interface~\cite{cni}~(CNI) plugin, which manages the lifetime of CXI services for all containers. CNI plugins interface with container runtime engines via a set of standardized commands, such as \ttt{ADD} and \ttt{DEL}, which respectively add networking components  to and remove them from the container's network namespace. 
These plugins are called while the container runtime is creating the container and before applications are launched inside the container.

The CXI CNI plugin is implemented as a \enquote{chained} plugin, meaning it modifies the network namespace configured by a previously called plugin. 
With this mode, we support the deployment of our CNI plugin in conjunction with any other CNI plugin, such as Flannel~\cite{flannel} or Cilium~\cite{cilium}. 

During \ttt{ADD}, our plugin (1) extracts the network namespace inode from the container under construction, (2) fetches a VNI from the VNI Service, and (3) creates a CXI service for the extracted netns inode and received VNI. 
If no VNI could be fetched from the VNI Service, the container will fail to launch.

During \ttt{DEL}, our plugin deletes any CXI service associated with the container being deleted. 
Automatic clean-up of unused CXI services and the associated VNI resources will therefore only occur if the CXI CNI plugin is available throughout the lifetime of all Slingshot-enabled containers.

Our CNI plugin only creates new CXI services if requested by the calling container via annotations. The request logic is outlined in Section~\ref{subsec:vni-service}. If a container does not request CXI communication capabilities, then our plugin performs no further operations and hence does not interfere with the container. 
To extract the container annotations and hence determine whether the container requests CXI capabilities, our CNI plugin queries the Kubernetes management plane.

\subsection{VNI Service}\label{subsec:vni-service}

\begin{figure}
    \centering
    \includegraphics[width=.95\linewidth]{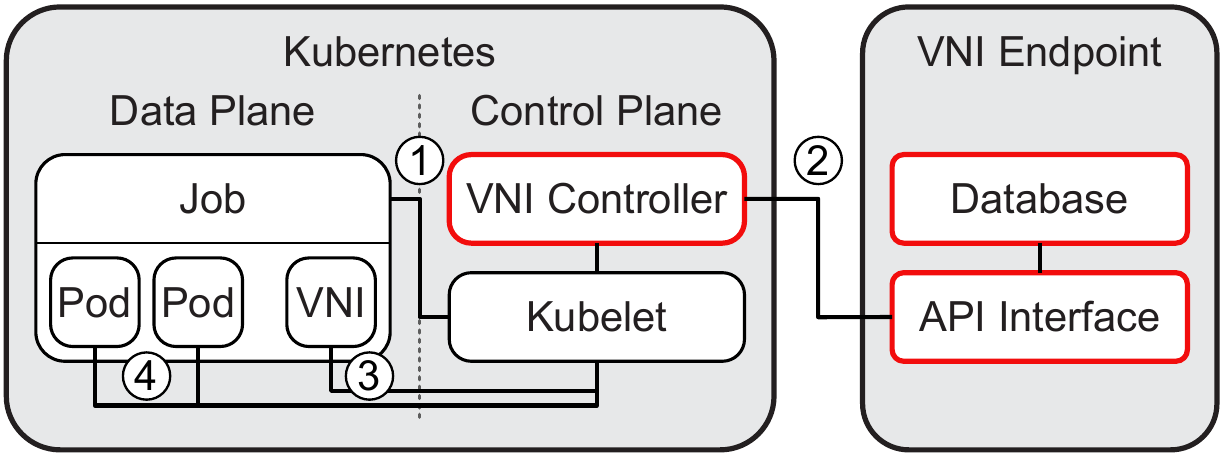}
    \caption{VNI Service, comprises VNI Endpoint and VNI Controller (red) | Numbers indicate logical flow of job admission.}
    \label{fig:vni-service}
\end{figure}

The Virtual Network Identifier (VNI) Service, labeled (C) in Figure~\ref{fig:cxi-vni-svc}, is a service running within Kubernetes that is responsible for managing the lifetime and association of Slingshot VNIs in a Kubernetes cluster. The VNI service consists of the \emph{VNI Endpoint}, which is a database and an accompanying API Interface, and the \emph{VNI Controller}, which integrates the VNI Endpoint into the Kubernetes cluster using a Custom Resource Definition. The VNI Endpoint is deployed as a pod in the cluster, while the VNI Controller resides in the cluster control plane.
Figure~\ref{fig:vni-service} outlines the architecture of the VNI Service. Numbers indicate the logical flow of job admission, which is described in more detail below.

\subsubsection{VNI Controller and Custom Resource}

Kubernetes provides Custom Resource Definitions (CRDs)~\cite{k8s_crd} to extend Kubernetes clusters with new functionality. We use a VNI CRD to express the cluster-wide VNI resource within Kubernetes and build a VNI controller based on Metacontroller~\cite{metacontroller} for resource management. 
An instance of the VNI CRD represents one VNI, which is an unsigned integer that represents the Virtual Network allocated on the cluster and configured on each involved CXI NIC. These VNI CRD instances are used to represent the assignment of VNIs to Kubernetes compute resources, such as pods or sets of pods. Kubernetes exposes various resources for managing a set of pods, such as Kubernetes Jobs, StatefulSets or ReplicaSets, all of which create a group of pods scheduled to run on one or more nodes. In this section, we will use Jobs as a representative for this class of resources. 

VNI CRDs reside in at most one Kubernetes namespace. We assume that users of a Kubernetes cluster are only permitted to operate on namespaces assigned to them. We further assume that users are not allowed to create or modify VNI CRD instances directly, leaving the VNI Controller as the only component that manages VNI CRDs.

\begin{figure}
    \centering
    \includegraphics[width=.95\linewidth]{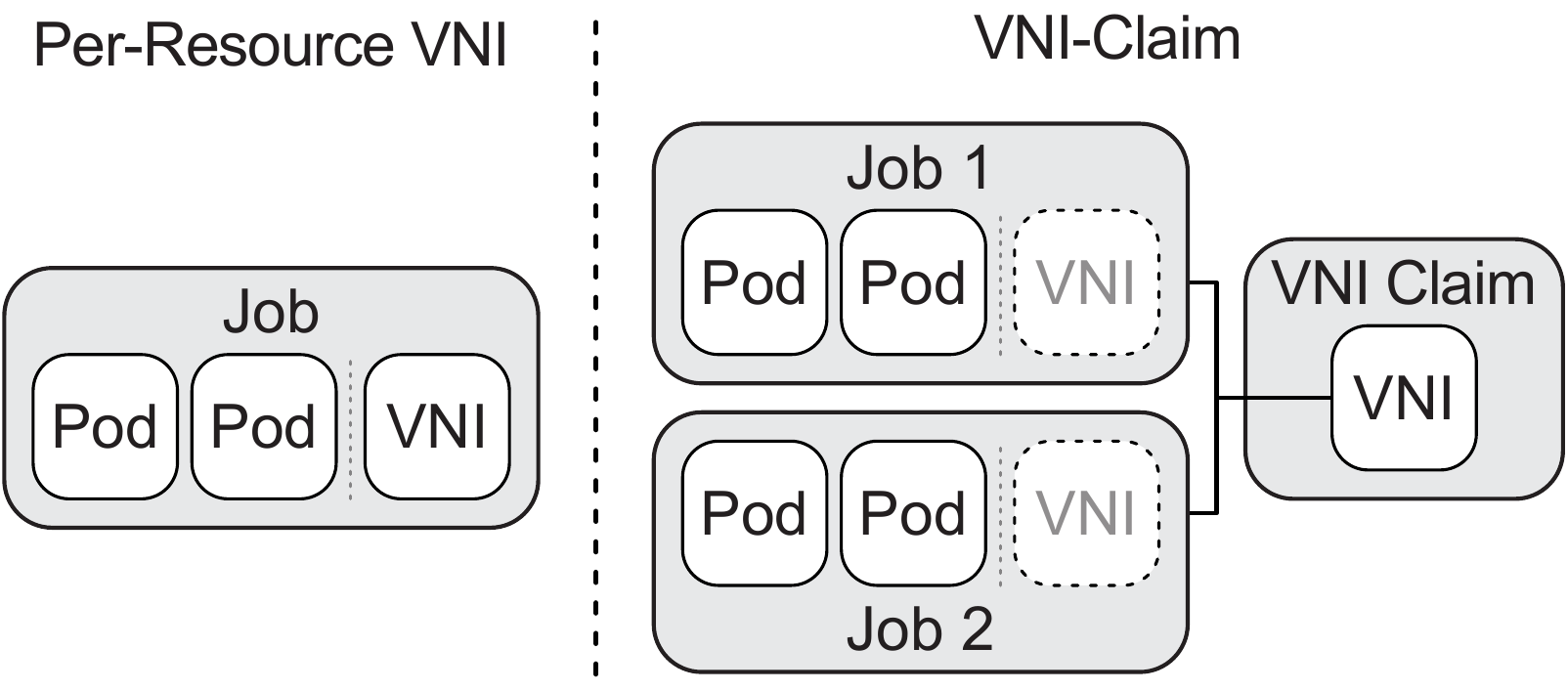}
    \caption{VNI Ownership Models | Per-Resource Model assigns each Job resource one VNI. VNI Claim Model assigns VNIs to VNI Claims, which multiple jobs can redeem.}
    \label{fig:vni-models}
\end{figure}

We provide two models for managing ownership and lifetime of VNI: \textit{Per-Resource VNIs}, and \textit{VNI Claims}. Figure~\ref{fig:vni-models} illustrates both models.

\textbf{Per-Resource VNIs}:~
The Per-Resource VNI model operates on the hierarchy level of job resources and associates ownership and lifetime of VNI CRD instances to individual job resource objects. Users can request Slingshot communication capabilities by adding the annotation \ttt{vni: true} to the job description. The VNI Controller will detect new jobs created with that annotation via an API call from the Kubelet (step (1) in Figure~\ref{fig:vni-service}), acquire a new VNI from the VNI Endpoint (2), and create a new VNI CRD instance representing the acquired VNI (3). We discuss the VNI acquisition in more detail below. When pods of the annotated job are created, the CXI CNI plugin fetches the acquired VNI from the VNI CRD instance. It establishes the respective CXI service (4), allowing the pods to use Slingshot within the virtual network identified by the acquired VNI.

In the Per-Resource VNI model, VNI CRD instances are entirely owned by the requesting job. As outlined in the previous Section, the CXI CNI plugin only launches pods if a VNI CRD instance has been created. Pods can therefore only launch when their acquisition request for a fresh VNI has been served. During job termination, the controller releases the VNI acquired from the VNI Service. However, job termination may precede pod termination. To avoid reusing still-active VNIs, we only hand out a VNI after it has been released for more than 30 seconds. We further require the termination grace period of pods not to exceed 30 seconds, which means that any straggling pod will be terminated at most 30 seconds after the owning job is terminated. The CXI CNI plugin enforces this termination grace period for all pods requesting a VNI.

Adding per-resource VNIs to existing workflows is straightforward, as it only requires an additional annotation. Listing~\ref{code:vni-job} shows an example Kubernetes job requesting a new VNI. 
A consequence of this ownership model is that pods can only communicate via Slingshot within the job owning the VNI. Pods of different jobs  therefore cannot communicate via Slingshot.

\begin{lstlisting}[style=yaml, caption=Example Kubernetes Job with Per-Resource VNI request, label=code:vni-job, float]
apiVersion: batch/v1
kind: Job
metadata:
  name: vni-test-job
  annotations:
    vni: true
spec:
  template:
    spec:
      containers:
        ...
\end{lstlisting}

\textbf{VNI Claims}:~
VNI Claims provide a way for several jobs and all associated pods to communicate with each other via Slingshot. We add VNI Claim as a second CRD, extending the VNI CRD. 
In the VNI Claim ownership model, we associate ownership and lifetime of VNI CRD instances with the VNI Claim object.

Before launching a set of new jobs, a user can create a VNI Claim instance, which is associated with a name that identifies the claim. Similar to VNI CRDs, VNI Claim CRDs are also namespaced, meaning that the user owning the namespace must ensure that the assigned VNI Claim name is unique within their namespace.
The VNI Controller will detect newly created VNI claims, acquire a new VNI through the VNI Endpoint, and create a new VNI CRD instance for that VNI claim.

To redeem a VNI claim inside a job, users can annotate their jobs with \ttt{vni: <vni-claim-name>}. The VNI Controller will search for a VNI claim that matches the annotation and attach the VNI to the job, if found. Similar to the Per-Resource VNI model, pods will eventually gain access to this VNI via the CXI CNI plugin. Jobs will fail to launch if no VNI claim with the annotated name has been found. Several jobs can be annotated with the same VNI claim name and will therefore be granted access to the same associated VNI, hence enabling Slingshot-based communication between jobs.

VNI claims must be created before job creation, meaning pods of a job can only launch if the requested VNI claim has already been made. To prevent users from deleting VNI claims while jobs are still using them, we track all jobs using a VNI claim and only allow VNI claim deletion if all users of that claim have terminated their jobs. This prevents handing out actively used VNIs to new jobs of potentially unrelated users.

Listing~\ref{code:vni-claim} shows an example VNI claim for name \ttt{vni-claim-test}, Listing~\ref{code:vni-job-claim} shows an example job using this claim.

\begin{lstlisting}[style=yaml, caption=Example VNI Claim, label=code:vni-claim, float]
apiVersion: v1
kind: VniClaim
metadata:
  name: vni-claim-test
  namespace: vnitest
spec:
  name: test
\end{lstlisting}

\begin{lstlisting}[style=yaml, caption=Example Kubernetes Job redeeming a VNI Claim, label=code:vni-job-claim, float]
apiVersion: batch/v1
kind: Job
metadata:
  name: vni-test-job
  namespace: vnitest
  annotations:
    vni: vni-claim-test
spec:
  template:
    spec:
      containers:
        ...
\end{lstlisting}

The VNI Controller is implemented as a Decorator Controller provided by Metacontroller. A Decorator Controller acts on already-created resources such as jobs, which match a specified pattern and \enquote{decorates} this resource with child objects. We configure the controller to listen to jobs annotated with the \ttt{vni} key and create VNI CRD instances as child objects. While the Metacontroller backend handles the actual creation of these child objects, the logic for how these objects should look is implemented in the VNI Endpoint, which our controller calls via webhooks.\\

\subsubsection{VNI Endpoint and Database}

Coordination of VNI assignments across all Kubernetes nodes is implemented via a database and an associated API endpoint. These components are deployed within the Kubernetes cluster as pods. 

\textbf{VNI Endpoint}:~
The VNI Endpoint serves as an interface between the VNI Controller and the VNI Database, which stores the ground truth for VNI assignments.

Metacontroller uses \enquote{apply semantics}: webhooks are called with information about an observed event, such as a newly created job, and expect the desired new cluster state as a response. This state is then applied to and reconciled with the cluster by the Metacontroller backend. Our VNI Endpoint exposes the endpoints \ttt{/sync} and \ttt{/finalize}. 

The \textbf{\ttt{/sync}} endpoint is called for both newly created jobs and VNI Claims. If the resource triggering the webhook should own the VNI, as is the case for jobs in the Per-Resource VNI model and VNI Claim CRD instances, then the endpoint will request a new VNI from the database and return the description of the new VNI CRD instance to the controller. The VNI Controller subsequently creates the VNI CRD object. If the triggering resource does not own the VNI, which is the case for jobs redeeming a VNI claim, then the endpoint will (1) search the database for the VNI associated with the claim, (2) add the triggering resource as a user to that VNI, and (3) return the description of a \enquote{virtual} or non-owning VNI CRD instance (dotted VNI object in Figure~\ref{fig:vni-models}, VNI Claim model). We use virtual VNI instances to maintain a one-to-one mapping between VNI CRD instances and job objects, and attach the VNI user removal logic to the deletion of the virtual VNI object. The \ttt{/sync} endpoint is idempotent, as it can be called for both update and creation events.

The \textbf{\ttt{/finalize}} endpoint is called for resources that are being deleted. For owning job resources, the owned VNI is released, and the VNI CRD is deleted. 
For non-owning job resources, the virtual VNI object is deleted, and the job is removed as a user of that VNI. 
For VNI Claims, the deletion request is only granted once all users of the VNI claim have been removed from the database, implying all pods using that VNI have terminated or are in the process of termination. VNI Claim deletion requests and subsequent release of the associated VNI will stall otherwise.

The VNI endpoint is responsible for managing the lifetime and association of VNIs, and the CXI CNI plugin enforces that a matching VNI must exist for jobs annotated with the \ttt{vni} annotation. Jobs annotated with that label will therefore only launch successfully if the VNI service is running.

\textbf{VNI Database}:~
The VNI Database stores all allocated VNIs and their associated users. In addition, we keep a log for all VNI allocation and release requests, as well as VNI user addition and removal requests. We use the relational database SQLite~\cite{sqlite} for storage.

Several operations performed on the database require multiple logical steps, such as VNI acquisition, which first checks for an available VNI and then inserts it into the allocation table if one is found.
Given the multi-threaded nature of the VNI Controller, race conditions or Time-of-Check to Time-of-Use errors can occur, for example, when two acquisition requests are received simultaneously.
We utilize the ACID properties of SQLite to prevent these errors by implementing all relevant database operations as atomic SQL transactions.

With the presented design and implementation, Kubernetes jobs can request and subsequently use Slingshot RDMA resources using a single job annotation. Access to these RDMA resources is granted on a per-job or job-set level, providing secure, per-tenant networking isolation.

\section{Evaluation}
\label{sec:evaluation}

In this section, we evaluate the performance overhead introduced by our Slingshot-Kubernetes integration. We evaluate two components: first, the RDMA communication overhead of applications using Slingshot within Kubernetes pods, and second, the job admission overhead. We assess our stack using a k3s-based~\cite{k3s} Kubernetes deployment running on two nodes of the OpenCUBE\cite{opencube} pilot system, which use Ampere Altra processors.
Table~\ref{tab:software} summarizes the versions of the involved software.
\begin{table}
    \centering
    \begin{tabular}{ll}
        \toprule
        \textbf{Software} & \textbf{Version}\\
        \midrule
            OpenSUSE & 15.5 \\
            k3s & v1.29.5 \\
            libfabric$^\dag$ & 2.1.0 \\
            Open MPI & 5.0.7 \\
            OSU Micro-Benchmarks & 7.3 \\
        \bottomrule
    \end{tabular}
    \caption{Overview of software versions used in experiment | Software marked with $^\dag$ has been patched to support the Slingshot-K8s integration.}
    \label{tab:software}
\end{table}

\subsection{Communication Overhead}

We measure communication overhead using the OSU Micro-Benchmark suite~\cite{omb}, which provides micro-benchmarks for individual MPI functions. Specifically, we measure point-to-point network throughput and latency using \ttt{osu_bw} and \ttt{osu_latency} respectively. We run all OSU benchmarks 10 times and use the default intra-benchmark iteration count of \num{10000} for bandwidth measurements and \num{20000} iterations for latency measurements. We use libfabric as the network abstraction library and apply patches to support our Slingshot integration, which exposes RDMA-based communication.

As a baseline, we measure both metrics on the host without involving Kubernetes. Next, we use the scheduling plugin Volcano~\cite{volcano} to run the MPI benchmarks within Kubernetes and specify \enquote{topology spread constraints}\footnote{Refer to \url{https://kubernetes.io/docs/concepts/scheduling-eviction/topology-spread-constraints/} for more information.} to spread the two involved containers onto the two nodes. We run the benchmarks within Kubernetes both with our Slingshot integration enabled (\ttt{vni:true}) and disabled (\ttt{vni:false}). Measurements of the MPI benchmark without our integration utilize a globally accessible VNI, which does not provide application-granular network isolation.

\begin{figure}
    \includegraphics[width=\linewidth]{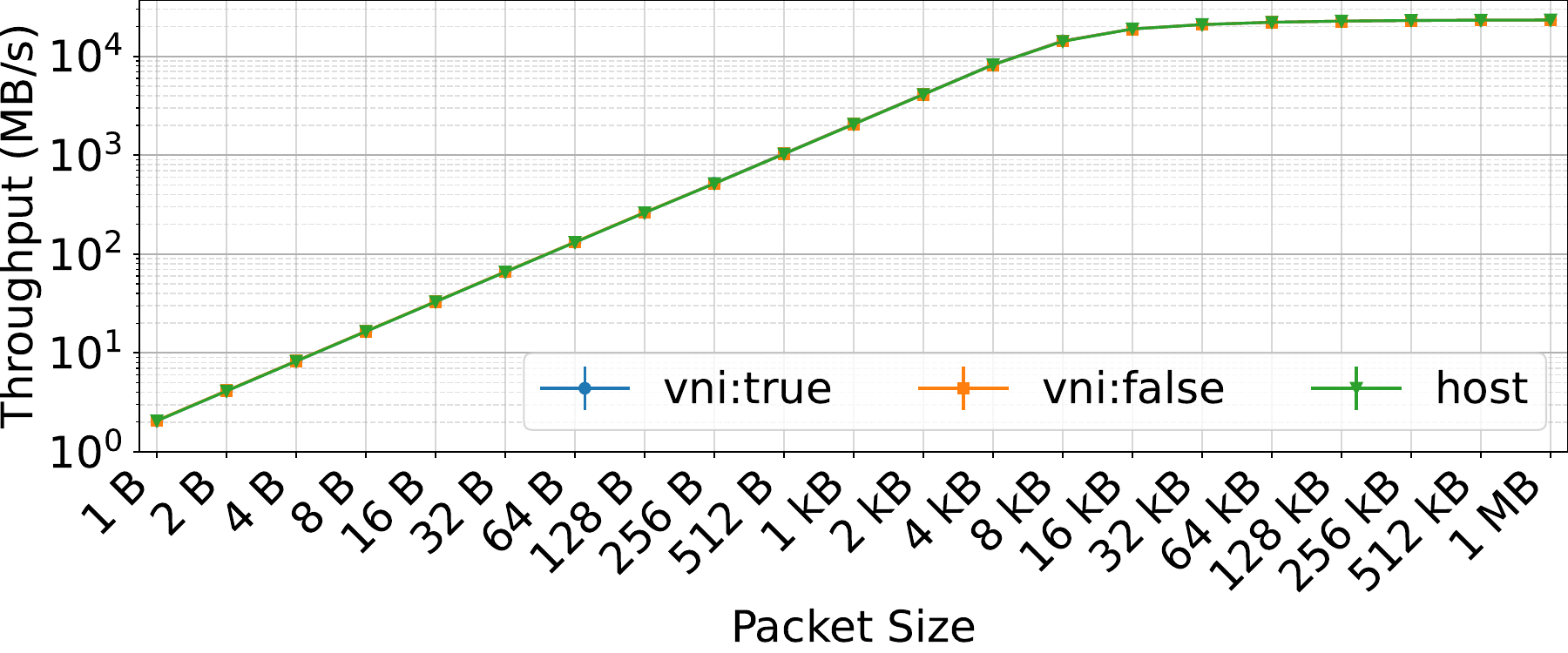}
    \caption{Average Throughput via \ttt{osu_bw} | Baseline run-to-run network jitter measured directly on the host in green. 10 iterations measuring 10k iterations of the respective MPI call.}
    \label{fig:raw_osu_bw}
\end{figure}

\begin{figure}
    \includegraphics[width=\linewidth]{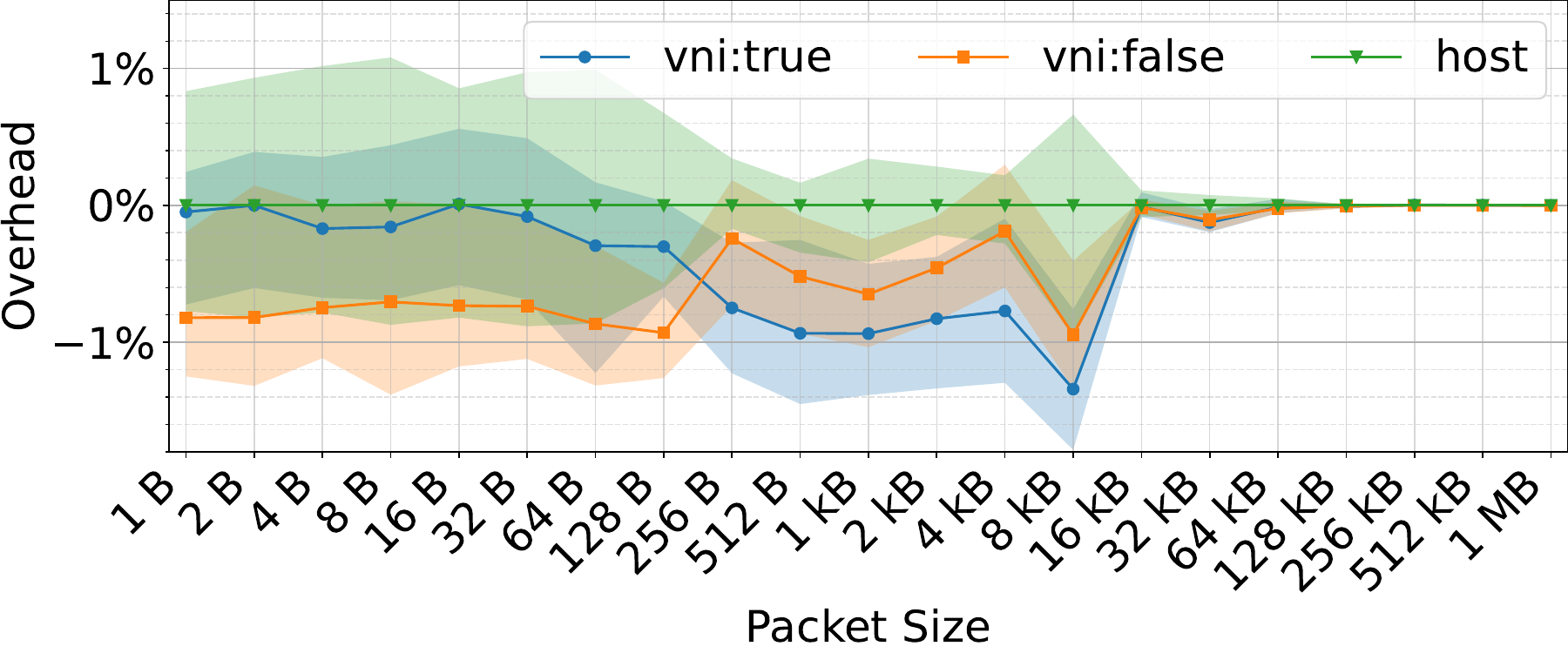}
    \caption{Average Throughput Overhead via \ttt{osu_bw} | Shaded regions indicate \qty{10}{\percent} and \qty{90}{\percent} percentile, baseline run-to-run network jitter measured directly on the host in green. 10 iterations measuring 20k iterations of the respective MPI call.}
    \label{fig:osu_bw}
\end{figure}

\begin{figure}
    \includegraphics[width=\linewidth]{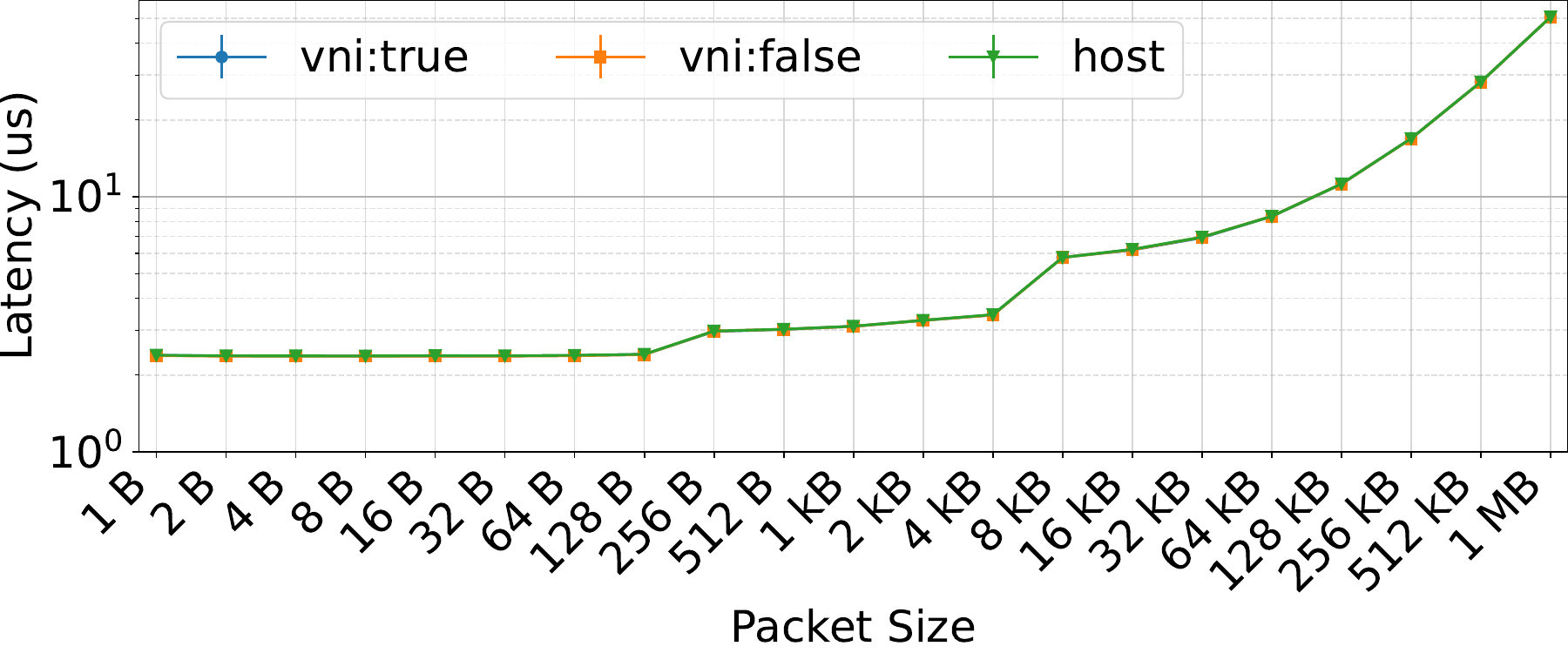}
    \caption{Average Latency via \ttt{osu_latency} | Baseline run-to-run network jitter measured directly on the host in green. 10 iterations measuring 20k iterations of the respective MPI call.}
    \label{fig:raw_osu_lat}
\end{figure}

\begin{figure}
    \includegraphics[width=\linewidth]{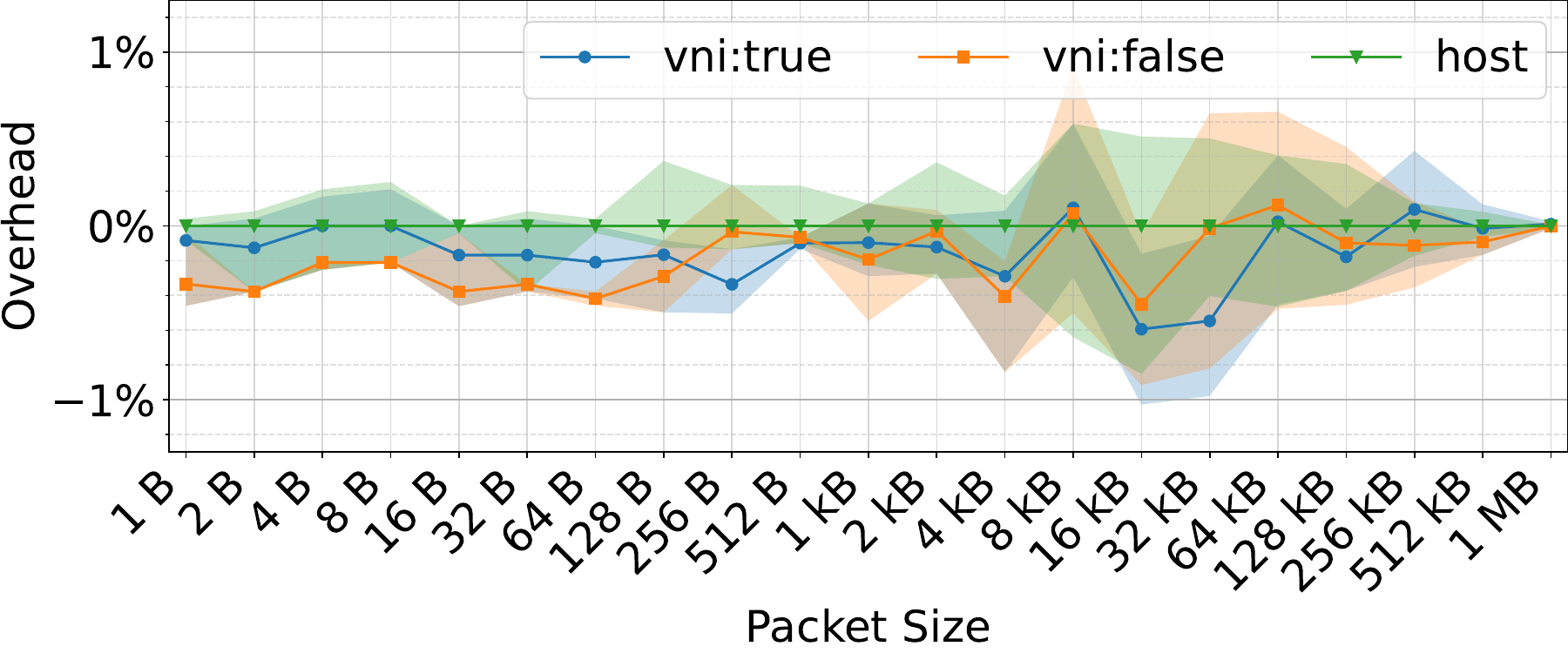}
    \caption{Average Latency overhead via \ttt{osu_latency} | Shaded regions indicate \qty{10}{\percent} and \qty{90}{\percent} percentile, baseline run-to-run network jitter measured directly on the host in green. 25 iterations measuring 20k iterations of the respective MPI call.}
    \label{fig:osu_lat}
\end{figure}

Figure~\ref{fig:raw_osu_bw} shows the average throughput measured using \ttt{osu_bw} of our integration (blue) against the throughput achieved without our integration (orange), as well as the baseline throughput achieved on the host (green). 
Figure~\ref{fig:osu_bw} shows the average throughput overhead. Shaded regions indicate \qty{10}{\percent} and \qty{90}{\percent} percentiles. 
Figure~\ref{fig:raw_osu_lat} shows the average latency measured using \ttt{osu_latency} and Figure~\ref{fig:osu_lat} shows the average latency overhead. 

The observed overhead is negligible and remains within \qty{1}{\percent}. We attribute it to inherent experimental variability rather than systematic overhead introduced by our integration.

\subsection{Job Admission Overhead}

We use the term \enquote{job admission delay} to refer to the time between job submission and the actual job start. Furthermore, we use the term \enquote{job admission overhead} to refer to the additional admission delay introduced by our Slingshot integration.
We measure job admission overhead using two approaches: (1) using a ramp-up ramp-down test, and (2) using a spike test. For both tests, we measure the time from job submission until completion. Jobs are configured to be deleted immediately after completion. Our measurements, therefore, include both job admission and deletion time, which encompass VNI allocation and release via the VNI Service, as well as CXI service creation and deletion via the CNI plugin. We launch jobs with one \ttt{alpine} container image each, running a single \ttt{echo} command. This minimizes the container image execution time in order to isolate the measurement of our figure of merit, namely, the job admission and deletion time. We pull the \ttt{alpine} image from a locally deployed harbor~\cite{harbor} container registry to minimize image pull time.

\subsubsection{Ramp Test}
The ramp test submits batches of $n$ jobs every second, where $n$ follows a ramp curve. During ramp-up, $n$ starts at one and increases to 10 with a step size of 1. During ramp-sustain, $n$ remains at 10 submitted jobs per second for 10 iterations. During ramp-down, $n$ decreases to 1 in steps of 1. This simulates scenarios where a large number of jobs is launched over a period of time instead of at once. We average results of $5$ measurement runs.

\begin{figure}
    \includegraphics[width=\linewidth]{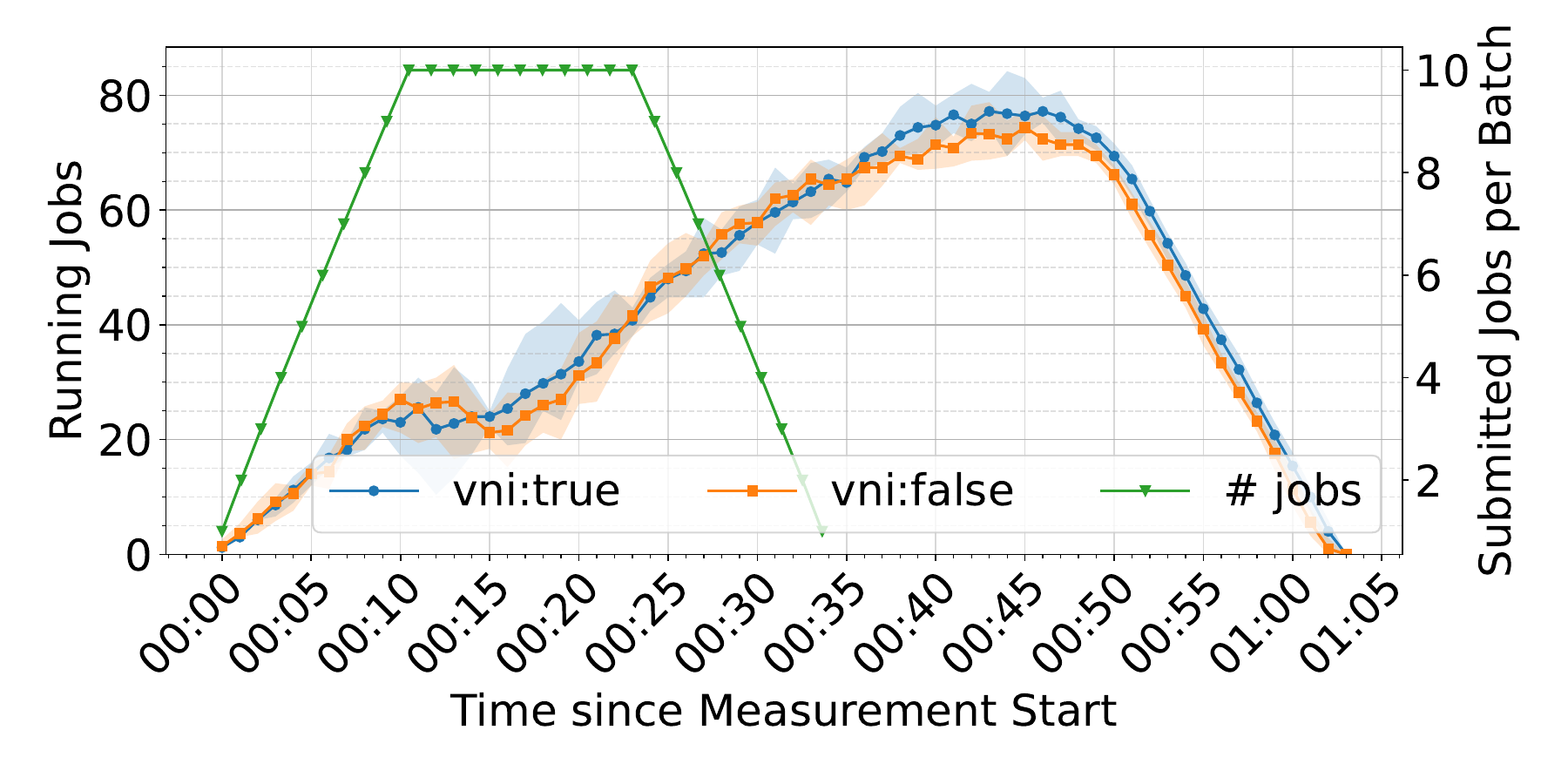}
    \caption{Number of actively Running Jobs during Ramp Test over time | Shaded areas indicate \qty{10}{\percent} and \qty{90}{\percent} percentiles across 5 runs. Number of submitted jobs per batch in green.}
    \label{fig:ramp:active_jobs}
\end{figure}

Figure~\ref{fig:ramp:active_jobs} shows the number of running jobs for each batch for the duration of the ramp test. The shaded area indicates the \qty{10}{\percent} and \qty{90}{\percent} percentiles across all runs. The blue and orange lines indicate runs with and without our Slingshot integration, respectively. The green line shows the number of jobs launched per job, following the ramp curve described above. Both runs with and without our stack show that job admission lags behind job submission, indicating that Kubernetes itself introduces a considerable job admission delay. Our integration shows that overhead is minimal and mostly confined to within run-to-run jitter.

\begin{figure}
    \includegraphics[width=\linewidth]{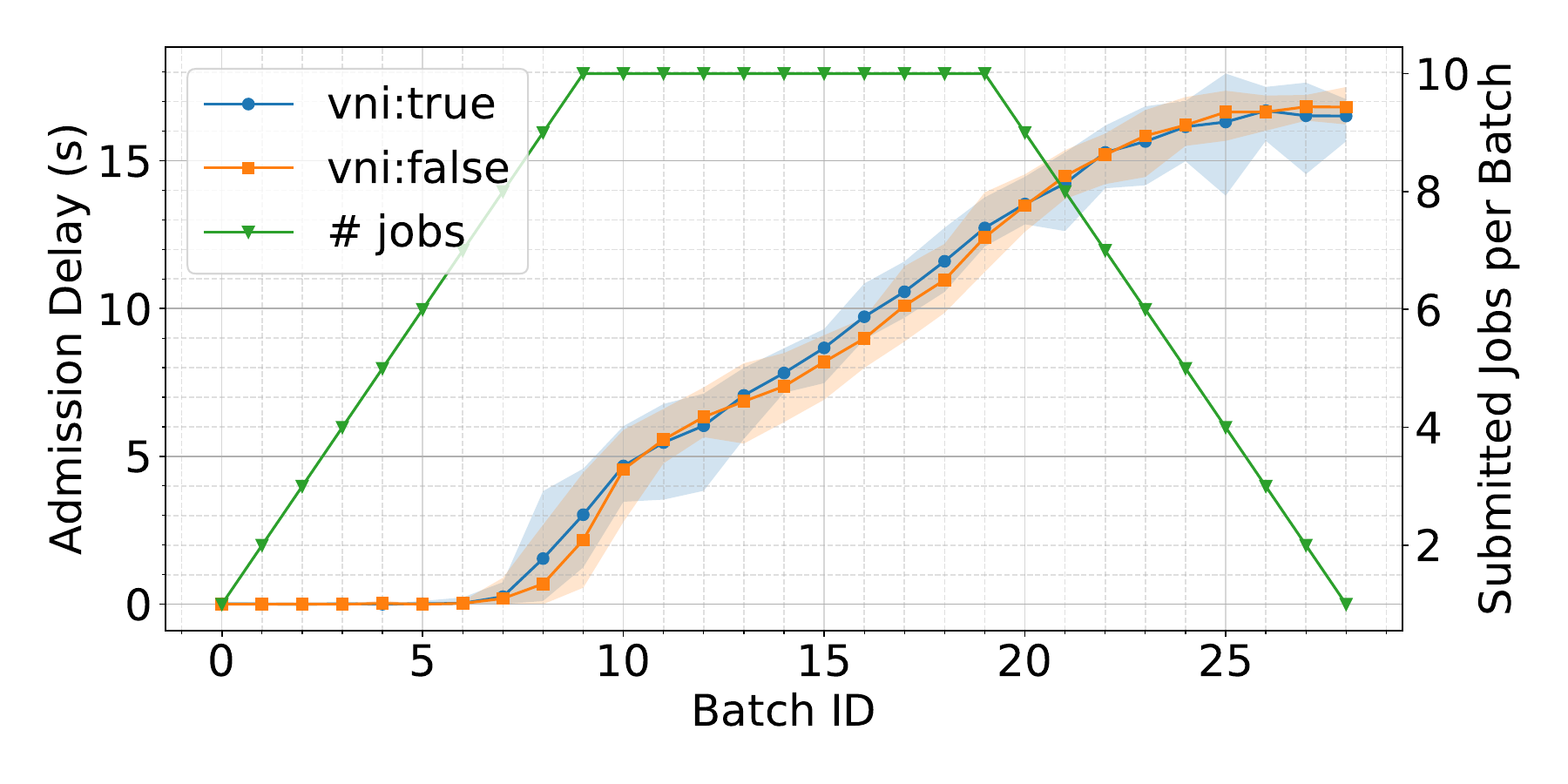}
    \caption{Job Admission Delay per Batch | Shaded areas indicate \qty{10}{\percent} and \qty{90}{\percent} percentiles across 5 runs. Number of submitted jobs per batch in green}
    \label{fig:ramp:job_delay_batch}
\end{figure}

Figure~\ref{fig:ramp:job_delay_batch} shows the job admission delay across submission batches. Both run types show job startup delay starts around batch 7 and increases with further batches. Overhead to the baseline measurement without our stack is minimal and confined to run-to-run jitter. We conclude that our integration does not impose significant overhead to job submission for ramp-up ramp-down load patterns.

\subsubsection{Spike Test}
The spike test submits $500$ jobs at once. This simulates maximum load on the system and reflects a worst-case load pattern. We average results of $5$ measurement runs.

\begin{figure}
    \includegraphics[width=\linewidth]{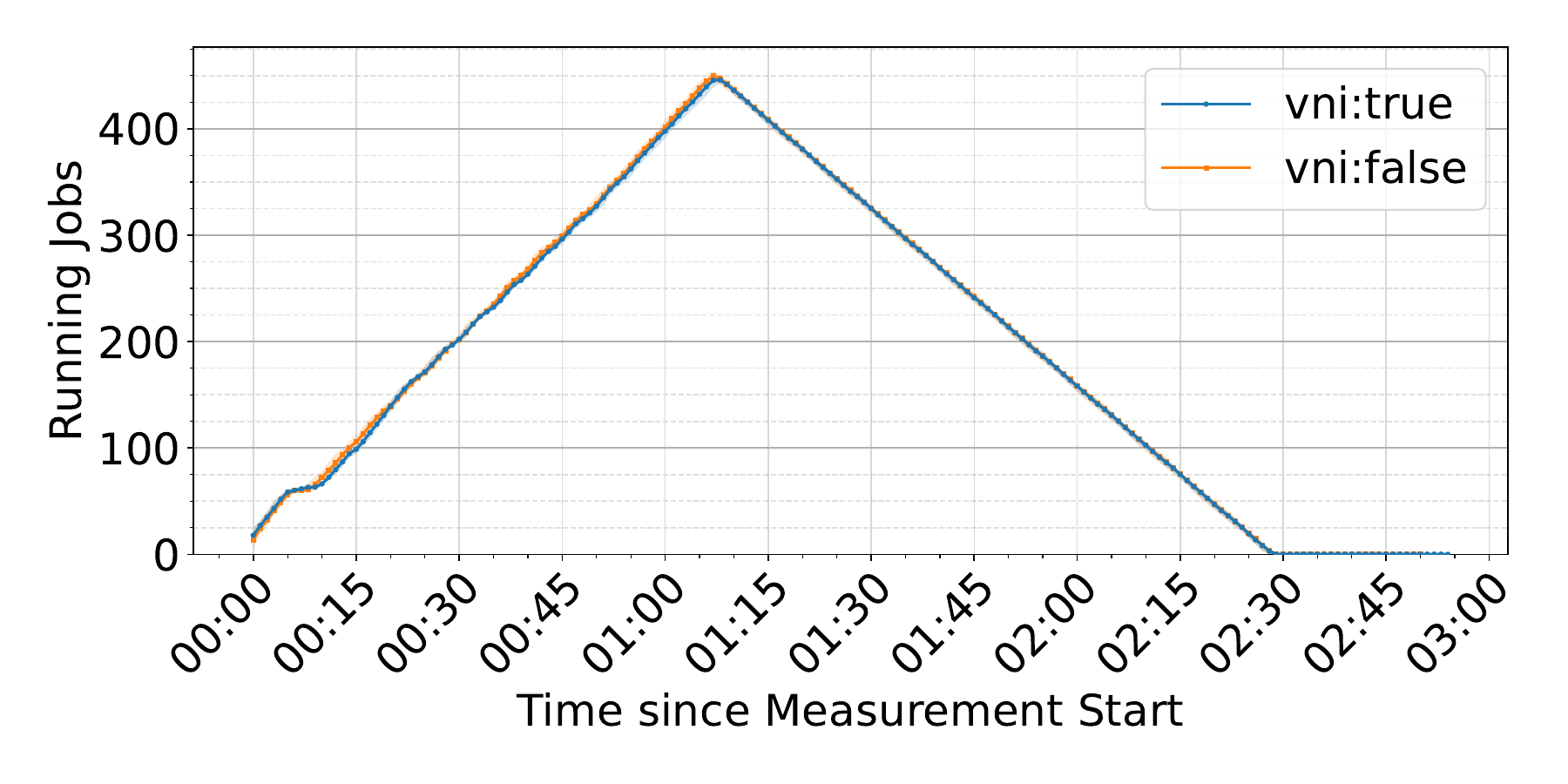}
    \caption{Number of actively Running Jobs during Spike Test over time | 500 jobs submitted at once. Shaded areas indicate \qty{10}{\percent} and \qty{90}{\percent} percentiles across 5 runs.}
    \label{fig:spike:active_jobs}
\end{figure}

Figure~\ref{fig:spike:active_jobs} shows the number of running jobs across the duration of the spike test. Both run types show that jobs are admitted and torn down linearly with a considerable delay. Our implementation shows no overhead compared to the baseline measurement, indicating that the majority of job admission overhead originates from the Kubernetes stack.

\begin{figure}
    \centering
    \begin{subfigure}{0.49\linewidth}
       \includegraphics[width=\linewidth]{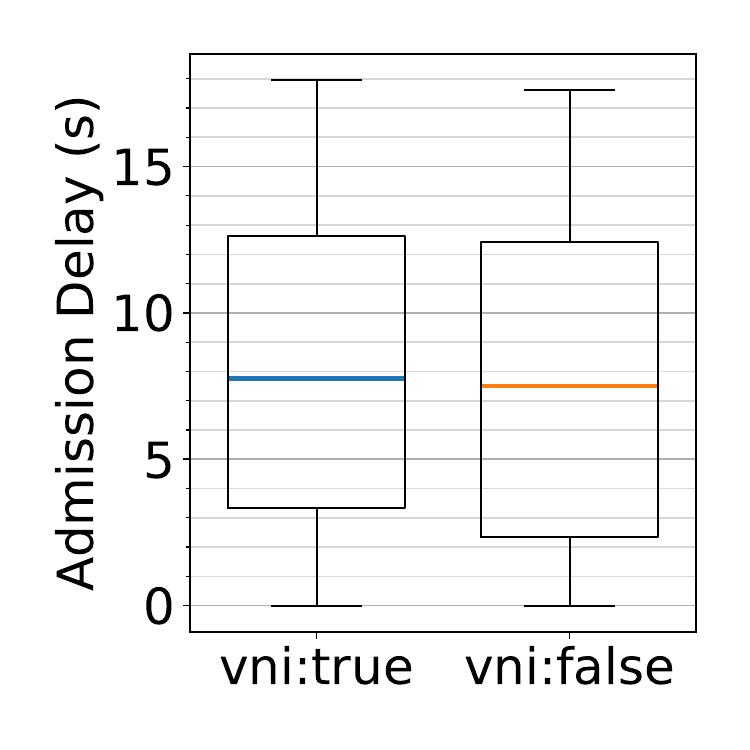}
        \caption{Ramp Test}
        \label{fig:ramp:job_delay}
    \end{subfigure}
    \hfill
    \begin{subfigure}{0.49\linewidth}
        \includegraphics[width=\linewidth]{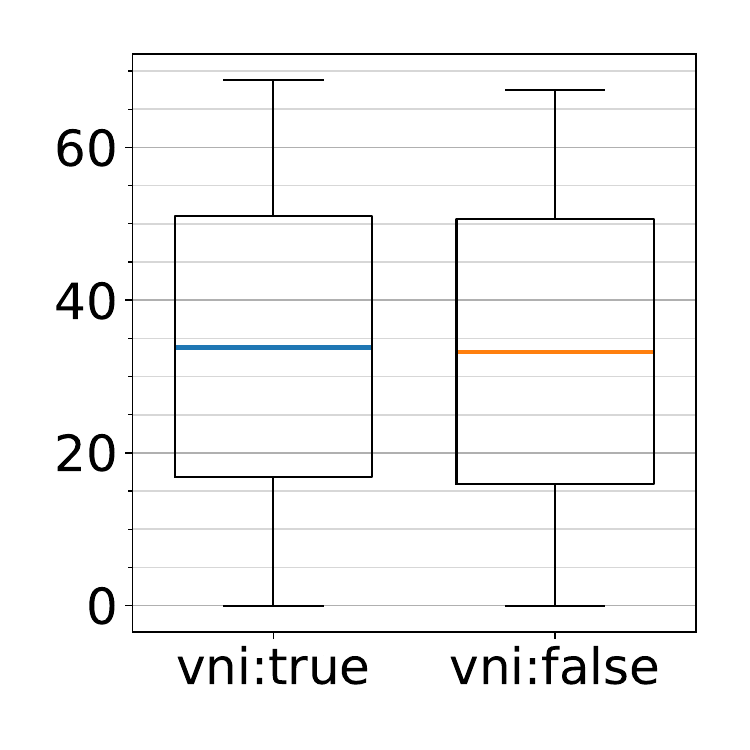}
        \caption{Spike Test}
        \label{fig:spike:job_delay}
    \end{subfigure}
\caption{Admission Delay for Ramp and Spike Test | Boxplots for admission delay in seconds over all individual jobs of all batches. }
\label{fig:job_delay}
\end{figure}

Figure~\ref{fig:job_delay} shows boxplots of the admission delay for both the ramp and spike tests over all involved jobs in all batches. The overall job admission overhead based on the admission delay median is \qty{3.5}{\percent} for the ramp test and \qty{1.6}{\percent} for the spike test.

In summary, we observe minimal job admission overhead of our integration for both load patterns and attribute the majority of job admission delay to the Kubernetes control plane.

\section{Related Work}\label{sec:related-work}

Xing et al. present Bedrock~\cite{Xing_Hsu_Qiu_Yang_Liu_Chen_2022}, which is a combined hardware- and software-stack providing secure RDMA networks. They utilize programmable network switches and eBPF-scripts to implement a variety of security mechanisms, such as RDMA source authentication and access control. In specific, Bedrock registers RDMA connections between communication parties and uses a P4-based program running on the network switch, which ensures that only packets from registered parties are routed. An eBPF-script running on the host uses the application PID to verify that only registered PIDs can access an RDMA queue pair. Bedrock targets InfiniBand-based RDMA networks.
The Rosetta Slingshot switch provides a similar feature to Bedrocks switch-level authentication, but operates on the level of VNIs. Bedrocks use of eBPF to keep track and authenticate users of RDMA queue pairs is conceptually similar to Slingshots use of CXI services, which also act as an authentication layer. Our work extends these CXI services to be compatible with container- or network-namespace-based workloads and integrates the authentication into Kubernetes, which has not been the focus of Bedrock.

Taranov et al. present sRDMA~\cite{Taranov_Rothenberger_Perrig_Hoefler_2020}, which extends the RDMA protocol with the transport type Secure Reliable Connection (SRC) Queue Pairs (QP), which provides source and data authentication as well as data encryption. The authors implement SRC using programmable NICs. sRDMA targets InfiniBand-based RDMA networks.
In sRDMA, all data sent using SRC QP is symmetrically encrypted using a key exchanged between the communicating parties during endpoint creation. This prevents malicious actors from illegitimately sending or receiving data, given that they are not in possession of the encryption keys. 
sRDMA solves confidentiality using encryption, which is a different approach to the one presented in this paper, namely extending the authentication or access layer to the RDMA network. Similar to Bedrock, integration into container-based workflows has not been the focus of sRDMA.

NVIDIA provides two variants of adding RDMA via InfiniBand to Kubernetes clusters: (1) via a shared RDMA device~\cite{k8s-rdma-shared-dev-plugin}, and (2) via Single-Root I/O Virtualization (SR-IOV)~\cite{ib-sriov-cni}. Variant~1 is conceptually similar to the approach presented in this paper, where access to RDMA NICs is provided to all requesting pods in a shared manner. Variant~2 uses SR-IOV functionality of InfiniBand NICs, which is currently not available for Slingshot NICs.
Both variants use a dedicated Kubernetes operator which handles resource management~\cite{network-operator}.

HPE provides an implementation for a Kubernetes device plugin for the Slingshot Cassini NIC~\cite{cxi-k8s-device-plugin}, which is still under development at the time of writing. The device plugin enables registration of CXI NICs as a resource in a Kubernetes cluster. During container creation, the plugin handles mounting the relevant CXI libraries and CXI character devices. In comparison to the implementation described in this paper, the device plugin does not handle CXI service management and instead assumes external management. In addition, these externally managed CXI services are not container-granular, which limits their use for providing secure, confidential communication via Slingshot for Kubernetes-based deployments.

\section{Conclusions}
\label{sec:conclusion}

In this paper, we introduced a novel integration of HPE Slingshot into Kubernetes, which enables secure, multi-tenant RDMA communication via Slingshot for converged HPC-Cloud systems.
We extended the Slingshot NIC driver and library with container support and developed a novel CNI plugin as well as a dedicated management service for integration into Kubernetes.
Based on our overhead evaluation, RDMA throughput and latency of workflows using our integration show no systematic overhead beyond run-to-run variation.
Job admission overhead of workflows in Kubernetes induced by our integration is minimal at around \qty{3.5}{\percent} and \qty{1.6}{\percent} for ramp-up-ramp-down, and spike load patterns, respectively. 

The source code for the introduced software stack, as well as measurement data and visualization software used in this paper, is available at: \url{https://github.com/caps-tum/paper-2025-shs-k8s}.

\bibliographystyle{IEEEtran}
\bibliography{bibliography}
\end{document}